\documentclass{statsoc}

\usepackage[a4paper]{geometry}

\usepackage{graphicx}
\usepackage[textwidth=8em,textsize=small]{todonotes}
\usepackage{amsmath}
\usepackage{natbib}

\usepackage{etoolbox}

\makeatletter
\patchcmd{\@makecaption}
  {\parbox}
  {\advance\@tempdima-\fontdimen2} 
  {}{}
\makeatother 

\usepackage{amsfonts}
\usepackage{amssymb}
\usepackage{booktabs} 
\usepackage{color}

\newcommand{\var}{\operatorname{Var}}

\title[Information monitoring of recurrent event endpoints]{Blinded continuous information monitoring of recurrent event endpoints with time trends in clinical trials}
\author{Tobias M{\"u}tze}
\address{Statistical Methodology, Novartis Pharma AG, Basel, Switzerland}
\email{tobias.muetze@novartis.com}
\author{Susanna Salem}
\address{Department of Medical Statistics, University Medical Center G{\"o}ttingen, G{\"o}ttingen, Germany.}
\author{Norbert Benda}
\address{Department of Medical Statistics, University Medical Center G{\"o}ttingen, G{\"o}ttingen, Germany.}
\address{Biostatistics and Special Pharmacokinetics Unit, Federal Institute for Drugs and Medical Devices, Bonn, Germany.}
\author{Heinz Schmidli}
\address{Statistical Methodology, Novartis Pharma AG, Basel, Switzerland}
\author[M{\"u}tze {\it et al.}]{Tim Friede}
\address{Department of Medical Statistics, University Medical Center G{\"o}ttingen, G{\"o}ttingen, Germany.}

\begin{document}
\begin{abstract}
Blinded sample size re-estimation and information monitoring based on blinded data has been suggested to mitigate risks due to planning uncertainties regarding nuisance parameters. Motivated by a randomized controlled trial in pediatric multiple sclerosis (MS), a continuous monitoring procedure for overdispersed count data was proposed recently. However, this procedure assumed constant event rates, an assumption often not met in practice. Here we extend the procedure to accommodate time trends in the event rates considering two blinded approaches: (a) the mixture approach modeling the number of events by a mixture of two negative binomial distributions, and (b) the lumping approach approximating the marginal distribution of the event counts by a negative binomial distribution. Through simulations the operating characteristics of the proposed procedures are investigated under decreasing event rates. We find that the type I error rate is not inflated relevantly by either of the monitoring procedures, with the exception of strong time dependencies where the procedure assuming constant rates exhibits some inflation. Furthermore, the procedure accommodating time trends has generally favorable power properties compared to the procedure based on constant rates which stops often too late. The proposed method is illustrated by the clinical trial in pediatric MS.
\end{abstract}

\section{Introduction}
\label{sec:intro}
Misspecification of nuisance parameters in the design of a clinical trial bears the risk of inconclusive results or wasteful use of resources when the variation in the data is larger or smaller than expected, respectively. To mitigate these risks, nuisance parameter based sample size re-estimation has been suggested and is commonly applied in clinical trials. Since re-estimation procedures based on blinded or non-comparative data generally lead to smaller bias and type I error rate inflation than unblinded procedures, and unblinded procedures bear the risk of compromising the integrity of the trial, these are preferred in regulatory guidance documents over unblinded approaches \citep*{EMAadaptive, FDAadaptive, EMAadaptive}.\par
The variability of the sample size resulting from a nuisance parameter based sample size re-estimation can be reduced by repeated estimation of the nuisance parameters during the course of the study. The power and expected sample size of designs with repeated re-estimation are similar to those of designs with a single re-estimation \citep*{friede2012blinded}. 
If taken to an extreme, repeated sample size re-estimation results in continuous monitoring with estimation of the nuisance parameters after every new data point. 
The trial is stopped once a sufficient level of information, i.e. precision of treatment effect estimate, is achieved.
This is analogous to event-driven trials in which the trial is stopped once a prespecified number of events is observed. \par
Blinded continuous monitoring was considered for normally distributed data by \cite{friede2012blinded} and was recently transferred to the setting of recurrent event data by \cite{friede2018blinded}. Whereas with normally distributed data only the variance needs to be monitored, with recurrent event data the information depends on the event rates, follow-up times, and overdispersion parameters, i.e., the between subject-variability. Without knowledge of the treatment groups the overall event rate can be estimated, which can then be split into group specific estimates under the assumption of a treatment effect hypothesized under the planning alternative. In randomized controlled trials, the follow-up times and the overdispersion parameters are usually assumed to be the same across treatment groups and therefore can fairly easily be estimated from blinded data pooled from all treatment groups. It could be shown that the application of such designs lead to shorter trial durations in comparison to traditional fixed designs while maintaining the power \citep*{friede2018blinded}.
For clinical trials with recurrent event data, continuous information monitoring differs from a repeated sample size re-estimation in that continuous information monitoring does not necessary result in a change in sample size. Depending on the duration of the recruitment period, continuous information monitoring can be materialized in changes in the study duration while keeping the total sample size as initially planned.
\par
The investigations by Friede et al. were motivated by a randomized controlled trial in pediatric multiple sclerosis where the annualized relapse rate was the primary endpoint \citep*{chitnis2018trial}. As this was the first large-scale double-blind, randomized controlled trial in this population, the design had to be based on adult data resulting in considerable uncertainty. Since the event rates were larger than assumed in the planning, the trial was stopped early based on results from blinded data looks. The analysis of this trial confirmed decreasing event rates over follow-up time, a trend previously also observed in a meta-analysis of adult data by \cite{nicholas2012time}. The procedure by \cite{friede2018blinded}, however, assumes constant event rates. Ignoring time trends in the event rates potentially may lead to biased estimates of the event rates as well as the overdispersion. Given the observed temporal trends in adult data, see \cite{nicholas2012time}, \cite{schneider2013blinded} had extended sample size re-estimation procedures for overdispersed count data, published by \cite{friede2010blinded,friede2010blinded1}, to account for these. To our knowledge, however, to date no blinded continuous monitoring procedure of the information for overdispersed count data is available accounting for time trends in the event rates. With this manuscript we want to close this gap.\par
The manuscript is organized as follows. In the following section some notation and underlying concepts are introduced before providing more background on the motivating clinical trial in pediatric multiple sclerosis in Section \ref{sec:example}. In Section \ref{sec:monitoring} blinded monitoring procedures are proposed and their operating characteristics are explored in a simulation study described in Section \ref{sec:simulations}. The motivating example is revisited in Section \ref{sec:example2}. We close with a brief discussion.
\section{Statistical model, hypothesis testing, and information}
\label{sec:statmodel}
In this section, we define a non-homogeneous Poisson process with Gamma frailty as a model for recurrent events with time trends.
Moreover, we define the statistical hypothesis of interest as well as an appropriate statistical test.
We conclude with defining information for the introduced statistical model. \par
Denote $N_{ij}(s)$ the number of events subject $j=1,\ldots, n_i$ in group $i=T,C$ has experienced up to study time $s$.
Study time $s$ is the time since randomization of a subject, i.e., the exposure time of a subject.
We assume that the recurrent events of each subject stem from a non-homogeneous Poisson process with a log-linear baseline rate $\exp(\alpha_{0} + \alpha_{1} s)$. 
Furthermore, proportionality of the rates between the treatment group and the control group is assumed. 
Therefore, the rate function for subject $j$ in group $i$, conditional on the subject-specific frailty $\nu_{ij}$, is given by
\begin{align}
\label{eq:model_trend}
\lambda_{ij}(s)|\nu_{ij}= \nu_{ij} \lambda_{i}(s) = \nu_{ij}\exp(\alpha_{0} + \alpha_{1} s) \exp(\beta x_i).
\end{align}
Here, $x_i$ is the group indicator which is zero in the control group, $x_C=0$, and one in the treatment group, $x_T=1$.
Moreover, the subject-specific frailty $\nu_{ij}$ is modeled as Gamma distributed, i.e. $\nu_{ij}\sim \Gamma(1/\phi, 1/\phi)$, with the Gamma distribution parameterized such that $\nu_{ij}$ has an expected value and a variance of 1 and $\phi$, respectively.
Hence, it follows that the number of events $N_{ij}(s)$ conditional on the frailty $\nu_{ij}$ are Poisson distributed, that is
\begin{align*}
N_{ij}(s)|\nu_{ij} \sim \operatorname{Pois}\left(\nu_{ij} \Lambda_{i}(s) \right),
\end{align*}
with the cumulative rate function
\begin{align*}
&\Lambda_{i}(s) = 
\displaystyle\int\limits_{0}^{s}\lambda_{i}(u)\,du =
\frac{1}{\alpha_{1}} \exp(\alpha_{0})\left(\exp(\alpha_{1}s)-1\right)\exp(\beta x_i) \\
\text{with} \qquad & 
\lambda_{i}(s) = \exp(\alpha_{0} + \alpha_{1} s) \exp(\beta x_i).
\end{align*}
The baseline cumulative rate function is denoted and given by $\Lambda_{C}(s)= \exp(\alpha_{0})\left(\exp(\alpha_{1}s)-1\right)/\alpha_{1}$.
It is important to emphasize that $\lambda_{i}(s)$  and $\Lambda_{i}(s)$ are not only functions of $s$ but also functions of $\alpha_{0}$, $\alpha_{1}$, and $\beta$. 
However, we do not explicitly mention the parameters $\alpha_{0}$, $\alpha_{1}$, and $\beta$ in our notation of $\lambda_{i}(\cdot)$ and $\Lambda_{i}(\cdot)$ for the sake of readability.
Marginally, the number of events $N_{ij}(s)$ follows a negative binomial distribution with rate $\Lambda_{i}(s)$ and shape parameter $\phi$ \citep*{ lawless1987negative, lawless1987regression}, that is
\begin{align*}
N_{ij}(s) \sim \operatorname{NegBin}\left(\Lambda_{i}(s), \phi \right).
\end{align*}
The expected value and the variance of the number of events $N_{ij}(s)$ are $\Lambda_{i}(s)$ and $\Lambda_{i}(s)(1+\phi \Lambda_{i}(s))$, respectively.
Thus, the variance increases in the shape parameter $\phi$. \par
In this manuscript, we are interested in the superiority of an experimental treatment over the control.
Under the assumption that smaller rates are better, superiority in the model above is given when the parameter $\beta$ is smaller than zero.
Thus, the question of superiority of the treatment over control can be written as the statistical testing problem
\begin{align*}
H_0: \beta \geq 0 \quad \text{vs.} \quad H_1: \beta < 0.
\end{align*}
To test the null hypothesis $H_0$, we employ a Wald test based on asymptotic maximum likelihood theory. 
Therefore, we discuss the maximum likelihood estimation of the parameters from the model above, $(\alpha_{0}, \alpha_{1}, \beta, \phi)$, and the parameters' asymptotic properties at first, followed by an introduction of the Wald test for $H_0$.
Let $S_{ij}^{(t)}$ be the exposure time of subject $j=1,\ldots, n_{i}$ in group $i=T,C$ at a calendar time $t$, and let $N_{ijt}=N_{ij}\left(S_{ij}^{(t)}\right)$ by the corresponding number of events. 
The study times, at which the events of a subject occurred, are denoted by $s_{ij1}, s_{ij2},\ldots$.
Then, according to \cite{lawless1987regression}, the likelihood function is given by
\begin{align}
\label{eq:likelihood}
\mathcal{L}(\alpha_{0}, \alpha_{1}, \beta, \phi|t) = 
\prod \limits_{i = T,C} \prod \limits_{j = 1}^{n_{i}} \left(\prod \limits_{k = 1}^{N_{ijt}}\frac{\lambda_{C}\left(s_{ijk}\right)}{\Lambda_{C}\left(S_{ij}^{(t)}\right)} \right) \times \frac{\Gamma(\phi^{-1} + N_{ijt})}{\Gamma(\phi^{-1})N_{ijt}!} \frac{\left(\phi\Lambda_{i}\left(S^{(t)}_{ij}\right)\right)^{N_{ijt}}}{\left(1 + \phi\Lambda_{i}\left(S^{(t)}_{ij}\right)\right)^{N_{ijt} + \phi^{-1}}}.
\end{align}
This results in the following log-likelihood function $\log\mathcal{L}(\cdot)$:
\begin{align}
\label{eq:loglikelihood}
\log\mathcal{L}(\alpha_{0}, \alpha_{1}, \beta, \phi|t) =& 
\sum_{i=T,C} \sum_{j=1}^{n_{i}} \sum_{k=1}^{N_{ijt}} [\alpha_{0} + \alpha_{1} s_{ijk}]  + 
\sum_{i=T,C} \sum_{j=1}^{n_{i}} N_{ijt}(\log(\phi) + x_{i}\beta) &\notag\\
&+ \sum_{i=T,C} \sum_{j=1}^{n_{i}} \left[\log\Gamma \left(N_{ijt} + \phi^{-1}\right) - \log\Gamma\left(\phi^{-1}\right) - \log\left(N_{ijt}!\right)\right] \notag\\
&- \sum_{i=T,C} \sum_{j=1}^{n_{i}} \left[(N_{ijt} + \phi^{-1}) \log \left( 1 + \phi\Lambda_{i}\left(S^{(t)}_{ij}\right)  \right) \right].
\end{align}
The maximum likelihood estimators $(\hat{\alpha}_{0t}, \hat{\alpha}_{1t}, \hat{\beta}_{t}, \hat{\phi}_{t})$ at calendar time $t$ are calculated by maximizing the log likelihood function \eqref{eq:loglikelihood}.
This maximization can be performed by finding the root of the system of equations $\nabla\log\mathcal{L}(\alpha_{0}, \alpha_{1}, \beta, \phi|t)=0$ using the Newton-Raphson method. 
We list the partial derivatives in Appendix \ref{appendix:ml_theory}.
The maximum likelihood estimators $(\hat{\alpha}_{0t}, \hat{\alpha}_{1t}, \hat{\beta}_{t}, \hat{\phi}_{t})$ are asymptotically normally distributed in the sense that 
\begin{align}
\label{eq:asymp_normal}
\sqrt{n}\left( 
\begin{pmatrix}
\hat{\alpha}_{0t} \\ \hat{\alpha}_{1t} \\ \hat{\beta}_{t} \\ \hat{\phi}_{t} 
\end{pmatrix}
-
\begin{pmatrix}
\alpha_{0} \\ \alpha_{1} \\ \beta \\ \phi 
\end{pmatrix}
\right)
\xrightarrow[n\to\infty]{\mathcal{D}}
\mathcal{N}\left(0, \Sigma\right).
\end{align}
Here, $\Sigma=\lim_{n\to \infty} n\mathbf{I}_{t}^{-1}$ with $\mathbf{I}_{t}\in \mathbb{R}^{4 \times 4}$ the Fisher information matrix which is a function of the sample size, the unknown parameter vector, and the individual exposure times at calendar time $t$.
For details, we refer to Appendix \ref{appendix:ml_theory}. 
Let $\mathbf{c}^{\prime} = \begin{pmatrix} 0 & 0 & 1 & 0 \end{pmatrix}$, we define the Wald statistic $T_{t}$ at calendar time $t$ by
\begin{align}
\label{eq:waldstat}
T_{t} =  \frac{\hat{\beta}_{t}}{\sqrt{\mathbf{c}^{\prime} \hat{\mathbf{I}}_{t}^{-1} \mathbf{c}}}
\end{align}
with $\hat{\mathbf{I}}_{t}$ the plug-in estimator of the Fisher information matrix obtained by plugging in the maximum likelihood parameter estimators into the formula of the Fisher information matrix. 
From the asymptotic normality of the maximum likelihood estimator follows that the Wald statistic $T_{t}$ is asymptotically standard normally distributed at the boundary of the parameter space defined by the null hypothesis $H_0$, that is $\beta=0$.
Therefore, the Wald test that rejects $H_0$ when $T_{t}$ is smaller than the $\alpha$-quantile $z_{\alpha}$ of a standard normal distribution is an asymptotic level $\alpha$ test for the null hypothesis $H_0$.\par
We conclude this section by recapitulating the concept of statistical information. 
In general, the information $\mathcal{I}$ for a treatment effect $\beta$ is the reciprocal of the variance of its estimator $\hat{\beta}$, that is $\mathcal{I} = 1 / \var(\hat{\beta})$ \citep*{jennison1999group}. 
The information $\mathcal{I}$ measures the knowledge about the unknown treatment effect $\beta$ with larger values corresponding to a smaller uncertainty about the unknown treatment effect. 
For the non-homogeneous Poisson process model with Gamma frailty, the information $\mathcal{I}_{t}$ at calendar time $t$ is given by 
\begin{align*}
\mathcal{I}_{t} 
= \frac{1}{\var(\hat{\beta}_{t})}
= \frac{1}{\mathbf{c}^{\prime} \mathbf{I}_{t}^{-1} \mathbf{c}}.
\end{align*}
Analogously to the Fisher information matrix $\mathbf{I}_{t}$, the information $\mathcal{I}_{t}$ is a function of the sample size, the parameter vector, and the individual exposure times. 
The information $\mathcal{I}_{t}$ increases when the sample size $n$, the individual exposure times, or the rates increase, and it decreases when the overdispersion parameter $\phi$ increases.
Since the information $\mathcal{I}_{t}$ is defined through the variance of the parameter estimate, the information is closely linked to the power of the previously introduced Wald test. 
For a parameter $\beta_{H_1}$ located in the parameter space of the alternative hypothesis $H_1$, the power increases as the information $\mathcal{I}_{t}$ increases. 
Moreover, for a given power $P$ and the significance level $\alpha$, the target information $\mathcal{I}_{fix}$ required for the Wald test to achieve a power $P$ for the parameter $\beta_{H_1}$ can be determined \citep*{jennison1999group,tsiatis2006information}:
\begin{align*}
\mathcal{I}_{fix} = \frac{(z_{1-\alpha} + z_{P})^2}{\beta_{H_1}^{2}}.
\end{align*}
It is worth noting that the target information $\mathcal{I}_{fix}$ only depends on the significance level $\alpha$, the target power $P$, and the assumed effect in the alternative $\beta_{H_1}$. 
When planning a clinical trial, the sample size $n$, the study duration, the accrual period, and the maximum individual exposure time are chosen such that the trial conveys the desired target information $\mathcal{I}_{fix}$ at the end of the trial for a parameter vector $(\alpha_{0}, \alpha_{1}, \beta_{H_1}, \phi)$. 
Unless the exposure times are identical for all subjects at the end of the trial, no closed form expression for converting the target information into the sample size, study duration, etc exists \citep*{schneider2013blinded}.
\section{Motivating example: Clinical trial in pediatric multiple sclerosis}
\label{sec:example}
Multiple sclerosis is a disease of the central nervous system that is in many patients characterized by periods of disease worsening, so-called relapses, followed by periods of recovery. 
\cite{chitnis2018trial} published results of a clinical trial (ClinicalTrials.gov Identifier: NCT01892722) assessing the efficacy and safety of fingolimod versus interferon beta-1a in pediatric multiple sclerosis.
The trial included 215 subjects which were randomized 1:1 between the two treatments. 
The primary endpoint was the annualized relapse rate determined by a negative binomial regression model of the number of relapses per subject. \par
The sample size of the clinical trial in pediatric multiple sclerosis was planned based on results of the TRANSFORMS clinical trial (ClinicalTrials.gov Identifier: NCT00340834) which assessed efficacy and safety of fingolimod in adults with relapsing-remitting multiple sclerosis, because no prior clinical trials in the pediatric population were available.
In detail, the sample size of the clinical trial in pediatric multiple sclerosis was planned assuming a relative reduction of 50\% in the annualized relapse rate from 0.36 for the interferon beta-1a arm to 0.18 for the fingolimod arm. 
With a fixed follow-up of two years per subject, a target power of 80\%, a two-sided significance level of 5\%, and an overdispersion parameter of $0.82$, a total sample size of 190 subjects was planned. 
In a blinded assessment of the accumulated information during the trial, it was determined that the trial would be overpowered when conducted as initially planned. 
Based on this blinded information assessment and in agreement with the regulatory agencies, the clinical trial design was changed from a fixed duration to a flexible duration and stopped early.\par
In the following, we analyze the relapses observed in the clinical trial in pediatric multiple sclerosis using the non-homogeneous Poisson model with a log-linear time trend introduced in Section \ref{sec:statmodel}.
Table \ref{table:example_fit} lists the parameter estimates for the standard negative binomial model, which does not account for time trends, and model \eqref{eq:model_trend} when applied to data from the pediatric multiple sclerosis trial. 
It is important to note that the primary analysis published by \cite{chitnis2018trial} was a negative binomial regression adjusted for treatment, region, number of relapses in the previous two years before study enrollment, and pubertal status. 
For illustrative purposes, we keep the model simple and do not adjust for covariates. \par
\begin{table}\small
\caption{{\label{table:example_fit}}Fit of negative binomial model (model without time trend) and model \eqref{eq:model_trend} (model with time trend) for data from the pediatric multiple sclerosis trial. The $95\%$ confidence intervals of the point estimates are shown in brackets. }
\centering
\begin{tabular}{{l}{c}{c}}
\toprule
\textbf{Parameter} & \textbf{Model without time trend} & \textbf{Model with time trend} \\ \midrule
$\hat{\alpha}_{0}$ & $-0.240\,[-0.521,0.042]$ & $-0.066\,[-0.431, 0.30]$\\
$\hat{\alpha}_{1}$ & - & $-0.236\,[-0.553, 0.081]$\\
$\exp\left(\hat{\beta}\right)$ &  $0.180\, [0.106, 0.305]$ & $0.184\, [0.109, 0.311]$\\
$\hat{\phi}$ & $1.31\, [0.58, 2.03]$ & $1.26\, [0.55, 1.97]$\\
\midrule
\textbf{Cumulative rates} & & \\
\midrule
$\hat{\Lambda}_{C}(2)$ & $1.57$ & $1.49$\\
$\hat{\Lambda}_{T}(2)$ & $0.283$ & $0.274$\\
\bottomrule
\end{tabular}
\end{table}
The estimated relative reduction of the relapse rate is $82\%$ for both models listed in Table \ref{table:example_fit} which is in accordance with the results published by \cite{chitnis2018trial}.
Comparing the models with and without time trend, the cumulative rates after two years, the estimated effect, and the estimates shape parameter are similar with relative differences of $5\%$ or less. 
Moreover, the estimated trend parameter is $\hat{\alpha}_{1}= -0.23637$. 
Thus, the estimated relapse rate decreases by $37.67\%$ within two years from $0.936$ to $0.584$ in the interferon beta-1a and from $0.172$ to $0.1$ in the fingolimod group.
We use this example in Section \ref{sec:simulations} to motivate the simulation setting and revisit the example in Section \ref{sec:example2}.
\section{Blinded continuous information monitoring for recurrent events with time trends}
\label{sec:monitoring}
\subsection{Basic setting and notation}
In this section we propose a procedure for blinded continuous information monitoring for the non-homogeneous Poisson model with a Gamma frailty introduced in Section \ref{sec:statmodel}. 
We start by outlining the concept of blinded continuous information monitoring and introduce notation related to the blinded sample. 
To begin with, the target information $\mathcal{I}_{fix}$ for the significance level $\alpha$, the power $P$, and the effect $\beta_{H_1}$ of interest is determined before the trial. 
Then, the clinical trial design, that is the sample size, study duration, accrual period, etc, are determined based on guesstimates for the nuisance parameters $\alpha_{0}$, $\alpha_{1}$, and $\phi$ such that the target information is reached at the end of the trial under the premise of correct nuisance parameter guesstimates.
However, instead of conducting the trial as initially designed, the information $\mathcal{I}_{t}$ is monitored continuously in the calendar time $t$ and the trial is stopped at the first point in time at which the monitored information exceeds the target information $\mathcal{I}_{fix}$.
After the trial is stopped, the null hypothesis $H_0$ is tested using the fixed sample Wald test introduced in Section \ref{sec:statmodel}.
It can occur that the information at the initially planned end of the trial is smaller than the target information. 
In this case, the trial could be continued beyond its initially planned duration to obtain the target information $\mathcal{I}_{fix}$.
Naturally, the question about how to continuously monitor the information arises. 
Here, we focus on information monitoring procedures that maintain blinding for reasons outlined in Section \ref{sec:intro}.
From a statistical perspective, the main challenge in designs with blinded continuous information monitoring is to find an estimator $\hat{\mathcal{I}}_{t}$ for $\mathcal{I}_{t}$ without knowing the treatment indicator of subjects.
The information $\mathcal{I}_{t}$ is a function of the nuisance parameters $\alpha_{0}$, $\alpha_{1}$, and $\phi$, the treatment effect $\beta$, and the individual exposure times $S_{ij}^{(t)}$.
Since data is blinded, $\beta$ cannot be estimated and we propose a continuous information monitoring procedure for the planning alternative $\beta_{H_1}$.
Furthermore, the Fisher information matrix $\mathbf{I}_{t}$, which is used to calculate the information $\mathcal{I}_{t}$, explicitly depends on the subject-specific treatment group indicator. 
Therefore, this section is split into two parts. 
In Section \ref{sec:monitoring:para_est} we propose two procedures for blinded estimation of the nuisance parameters. 
In Section \ref{sec:monitoring:info} we illustrate how to estimate the Fisher information matrix, and therefore the information, without knowing the treatment group indicator. \par
For the blinded data, the notation introduced in Section \ref{sec:statmodel} is changed by substituting the index $i$ by $(b)$ and the upper limit of index $j$ is changed from $n_{i}$ to $m$.
In detail, at calendar time $t$, subject $j=1,\ldots, m$ has an exposure time of $S_{(b)j}^{(t)}$ and has experienced $N_{(b)jt}$ events at study times $s_{(b)jk}$ with $k=1,2,\ldots$.
Let $w_{i} = n_{i}/n$ be the proportion of patients to be randomized into group $i=T,C$ which is assumed to be known. 
\subsection{Blinded estimation of nuisance parameters}
\label{sec:monitoring:para_est}
\subsubsection{Mixture approach}
\label{sec:blindmixture}
In a randomized trial with $w_{i}, i=T,C$, a subject $j=1,\ldots, m$ from the blinded sample has with probability $w_{T}$ been randomized to the treatment group and with probability $w_{C}$ to the control group. 
Thus, the cumulative number of events $N_{(b)jt}$ from a subject $j$ in the blinded sample follows a mixture of two negative binomial distributions, that is 
\begin{align}
\label{eq:distmix}
N_{(b)jt} \sim w_{T}\operatorname{NegBin}\left(\Lambda_{T}(S_{(b)j}^{(t)}), \phi \right) + w_{C}\operatorname{NegBin}\left(\Lambda_{C}(S_{(b)j}^{(t)}), \phi \right).
\end{align}
Modeling the blinded sample through a mixture of two distributions has also been considered by \cite{asendorf2017modelling,asendorf2018sample} in the context of longitudinal count data.
Since we aim to only estimate the nuisance parameters, we replace the cumulative rate function $\Lambda_{T}(s)$ in \eqref{eq:distmix} under the assumption of a treatment effect $\beta_{H_1}$ by $\Lambda_{C}(s)\exp(\beta_{H_1})$.
From \eqref{eq:distmix} it follows that the log-likelihood of the blinded sample is given by
\begin{align}
\label{eq:loglikelihoodmix}
&\log\mathcal{L}_{mix}(\alpha_{0}, \alpha_{1}, \phi|t) \\ 
=& 
\sum_{j=1}^{m} \sum_{k=1}^{N_{(b)jt}} [\alpha_{0} + \alpha_{1} s_{(b)jk}]  + 
\sum_{j=1}^{m} N_{(b)jt}\log(\phi) &\notag\\
&+ \sum_{j=1}^{m} \left[\log\Gamma \left(N_{(b)jt} + \phi^{-1}\right) - \log\Gamma\left(\phi^{-1}\right) - \log\left(N_{(b)jt}!\right)\right] \notag\\
&- \sum_{j=1}^{m} \log\left(\frac{w_{T}\exp\left(\beta_{H_1}\right)^{N_{(b)jt}}}{\left(1+\phi \Lambda_{C}\left(S_{(b)j}^{(t)}\right)\exp\left(\beta_{H_1}\right)\right)^{N_{(b)jt}+\phi^{-1}}} + \frac{w_{C}}{\left(1+\phi \Lambda_{C}\left(S_{(b)j}^{(t)}\right)\right)^{N_{(b)jt}+\phi^{-1}}}\right).
\end{align}
Then, the maximum likelihood estimator of the nuisance parameters $\alpha_{0}$, $\alpha_{1}$, and $\phi$ at calendar time $t$ are defined by 
\begin{align}
\label{eq:blindestmix}
\left(\hat{\alpha}_{0(b)t}, \hat{\alpha}_{1(b)t}, \hat{\phi}_{(b)t}\right):= \arg\max\limits_{\alpha_{0}, \alpha_{1}, \phi} \log\mathcal{L}_{mix}(\alpha_{0}, \alpha_{1}, \phi|t) .
\end{align}
It is important to emphasize that the mixture approach for blinded estimation of the nuisance parameters through \eqref{eq:blindestmix} differs from expectation-maximization (EM) algorithm-based procedures for blinded parameter estimation in that the EM algorithm-based procedures also estimate the treatment effect. 
The appropriateness of EM algorithm-based procedures has been controversially discussed in the past \citep*{friede2002inappropriateness,waksman2007assessment,cook2009two,schneider2013robustness,cook2013authors}.
\subsubsection{Lumping approach}
The lumping approach for blinded nuisance parameter estimation in a non-homogeneous Poisson model with Gamma frailty was proposed by \cite{schneider2013blinded}, who extended a previous proposal by \cite{friede2010blinded} to dependent event rates.
The idea is to approximate the marginal distribution of the number of events $N_{(b)jt}$ by a negative binomial distribution, that is
\begin{align}
\label{eq:distlump}
N_{(b)jt} \mathrel{\dot{\sim}} \operatorname{NegBin}\left(\Lambda_{(b)}\left(S_{(b)j}^{(t)}\right), \phi\right),
\end{align}
with the cumulative rate function $\Lambda_{(b)}(\cdot)$ given by 
\begin{align*}
  \Lambda_{(b)}(s) 
= w_{T}\Lambda_{C}(s)\exp(\beta_{H_1}) + w_{C} \Lambda_{C}(s)
=  \frac{1}{\alpha_{1}}\exp(\alpha_{0})\left(\exp(\alpha_{1}s) - 1\right)  \left( w_{T}\exp(\beta_{H_1}) + w_{C} \right).
\end{align*}
As the last display shows, the cumulative rate function $\Lambda_{(b)}(\cdot)$ of the blinded sample is modeled by a mixture of the cumulative rate function from the treatment arm and the control arm with the weight of each part equal to the proportion of the sample size allocated to the respective arm. 
The blinded estimators for the nuisance parameters $\alpha_{0}$, $\alpha_{1}$, and $\phi$ at calendar time $t$ are obtained by maximizing the likelihood function $\mathcal{L}_{(b)}(\alpha_{0}, \alpha_{1}, \phi|t)$ of the blinded sample, 
\begin{align}
\label{eq:blindestlump}
\left(\hat{\alpha}_{0(b)t}, \hat{\alpha}_{1(b)t}, \hat{\phi}_{(b)t}\right)= \arg\max\limits_{\alpha_{0}, \alpha_{1}, \phi} \mathcal{L}_{(b)}(\alpha_{0}, \alpha_{1}, \phi|t).
\end{align}
The likelihood function $\mathcal{L}_{(b)}(\alpha_{0}, \alpha_{1}, \phi|t)$ of the blinded sample is the likelihood of a sample of independent negative binomial distributed random variables with rate parameter and dispersion parameter as in \eqref{eq:distlump}.\par
The blinded maximum likelihood estimates \eqref{eq:blindestlump} are not consistent under model \eqref{eq:distmix}. 
In particular, the parameter estimator $\hat{\phi}_{(b)}$ overestimates the dispersion parameter $\phi$ as it accounts for the within-group and the between-group variability in model \eqref{eq:distlump}.
For the lumping approach, we modeled the blinded data set by a negative binomial distribution. 
Technically, the assumption of a single negative binomial distribution is incorrect as the blinded sample follows a mixture of two negative binomial distribution as illustrated in Section \ref{sec:blindmixture}.
However, as previous research in the context of blinded sample size adjustments for negative binomial data has shown, the lumping approach is appropriate for blinded nuisance parameter estimation \citep*{schneider2013blinded, friede2010blinded, schneider2013robustness}.
\subsection{Blinded estimation of information}
\label{sec:monitoring:info}
The entries of the Fisher information matrix $\mathbf{I}_{t}$ relevant for calculating information $\mathcal{I}_{t}$ are sums of subject-specific values $f_{i}(S_{ij}^{(t)}, \alpha_{0}, \alpha_{1}, \beta, \phi)$ over $j=1,\ldots, n_{i}$ with either $i=T,C$ or $i=T$. 
Thus, the entries of $\mathbf{I}_{t}$ have one of the following structures
\begin{align}
\label{eq:genericsum1}
& \sum_{i=T,C} \sum_{j=1}^{n_{i}} f_{i}\left(S_{ij}^{(t)}, \alpha_{0}, \alpha_{1}, \beta, \phi\right),\\
\label{eq:genericsum2}
& \sum_{j=1}^{n_{T}} f_{T}\left(S_{Tj}^{(t)}, \alpha_{0}, \alpha_{1}, \beta, \phi\right).
\end{align}
For instance, the entry of $\mathbf{I}_{t}$ which corresponds to the negative second partial derivative with respect to $\alpha_{0}$ is the sum of 
\begin{align*}
f_{i}\left(S_{ij}^{(t)}, \alpha_{0}, \alpha_{1}, \beta, \phi\right) = \frac{\Lambda_{i}\left(S_{ij}^{(t)}\right)}{1 + \phi\Lambda_{i}\left(S_{ij}^{(t)}\right)}
\end{align*}
over $j=1,\ldots, n_{i}$ and $i=T,C$.
The summands $f_{i}(\cdot)$ are not identical between the two groups, that is $f_{T}(\cdot)\neq f_{C}(\cdot)$,  under the alternative.
Therefore, a blinded estimator of the Fisher information $\mathbf{I}_{t}$ is not obtained by simply plugging in the blind estimator for the parameters $\alpha_{0}$, $\alpha_{1}$, and $\phi$ from Section \ref{sec:monitoring:para_est} into the sums.
However, when the exposure times in both treatment groups have the same distribution at a given calendar time $t$, the exposure times $S_{(b)j}^{(t)}$ from the blinded sample at calendar time $t$ are distributed as the exposure times in each treatment group. 
This results in the following approximation of the sums \eqref{eq:genericsum1} and \eqref{eq:genericsum2} by sums utilizing the exposure times from the blinded sample:
\begin{align}
\label{eq:blindsum1}
& \sum_{i=T,C} \sum_{j=1}^{n_{i}} f_{i}\left(S_{ij}^{(t)}, \alpha_{0}, \alpha_{1}, \beta, \phi\right) \approx 
 \sum_{i=T,C} \sum_{j=1}^{m} w_{i} f_{i}\left(S_{(b)j}^{(t)}, \alpha_{0}, \alpha_{1}, \beta, \phi\right) ,\\
\label{eq:blindsum2}
& \sum_{j=1}^{n_{T}} f_{T}\left(S_{Tj}^{(t)}, \alpha_{0}, \alpha_{1}, \beta, \phi\right) \approx
 \sum_{j=1}^{m} w_{T}f_{T}\left(S_{(b)j}^{(t)}, \alpha_{0}, \alpha_{1}, \beta, \phi\right).
\end{align}
The sums on the right side of \eqref{eq:blindsum1} and \eqref{eq:blindsum2} only use information about the exposure times that is available in the blinded sample. 
Therefore, a blinded estimator $\hat{\mathbf{I}}_{(b)t}$ of the Fisher information matrix $\mathbf{I}_{t}$ at calendar time $t$ is obtained in two steps. 
Firstly, sums that make up the entries of the Fisher information matrix are rewritten as illustrated in \eqref{eq:blindsum1} and \eqref{eq:blindsum2}. 
Secondly, the nuisance parameters $\alpha_{0}$, $\alpha_{1}$, and $\phi$ are replaced by blinded estimates, which we proposed in Section \ref{sec:monitoring:para_est}, and the effect $\beta$ is replaced by the planning alternative $\beta_{H_1}$.\par
We denote the resulting blinded continuous information monitoring procedure by \textit{BCM-Trend--Lump} if the nuisance parameters are estimated based on the lumping approach and by \textit{BCM-Trend--Mix} if the nuisance parameters are estimated based on the mixture approach. 
\section{Simulation study}
\label{sec:simulations}
\subsection{Purpose of simulation study and motivation of scenarios}
In this section we assess the operating characteristic of the proposed blinded continuous information monitoring procedure. 
The focus is on two settings. 
Firstly, when a sponsor plans a clinical trial, major design aspects are the sample size and the corresponding power of the clinical trial under the assumed effect $\beta_{H_1}$. 
Thus, an important requirement on blinded continuous information monitoring procedures is that the trial's target power $P$ is maintained when the sample size planning assumptions are fulfilled. 
We refer to the setting, for which the planning assumptions are fulfilled, as \textit{Setting I}.
A sponsor's motivation for conducting a clinical trial with a blinded continuous information monitoring procedure is to stop the trial early, that is to reduce the sample size or the study duration, while maintaining the target power $P$ when the initially planned trial is overpowered. 
Therefore, in \textit{Setting II}, the  clinical trials are overpowered due to misspecified planning assumptions.
In both settings, we are interested in multiple performance measures. 
Firstly, we assess the type I error rate of the blinded continuous information monitoring procedure since a regulatory requirement of adaptive designs is to control the type I error rate \citep*{FDAadaptive, EMAadaptive}.
Additionally, motivated by regulatory requirements, we evaluate the bias of the treatment effect estimator for designs with blinded continuous information monitoring.
From the sponsor's perspective, the power as well as the distributions of the study duration and the sample size of designs with information monitoring are of interest. \par
\begin{figure}[ht]
\centering
\includegraphics[width=0.49\textwidth]{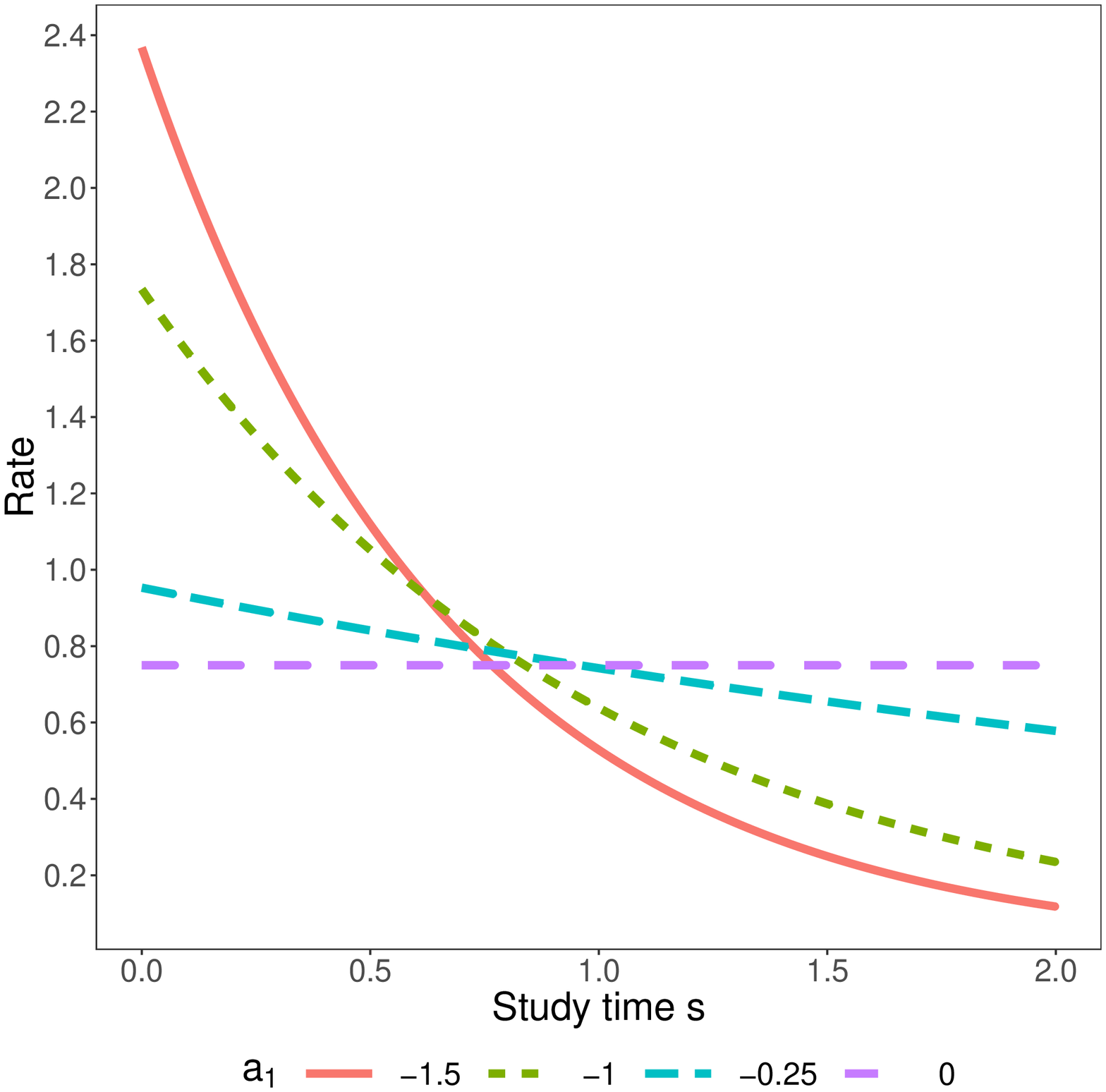}
\includegraphics[width=0.49\textwidth]{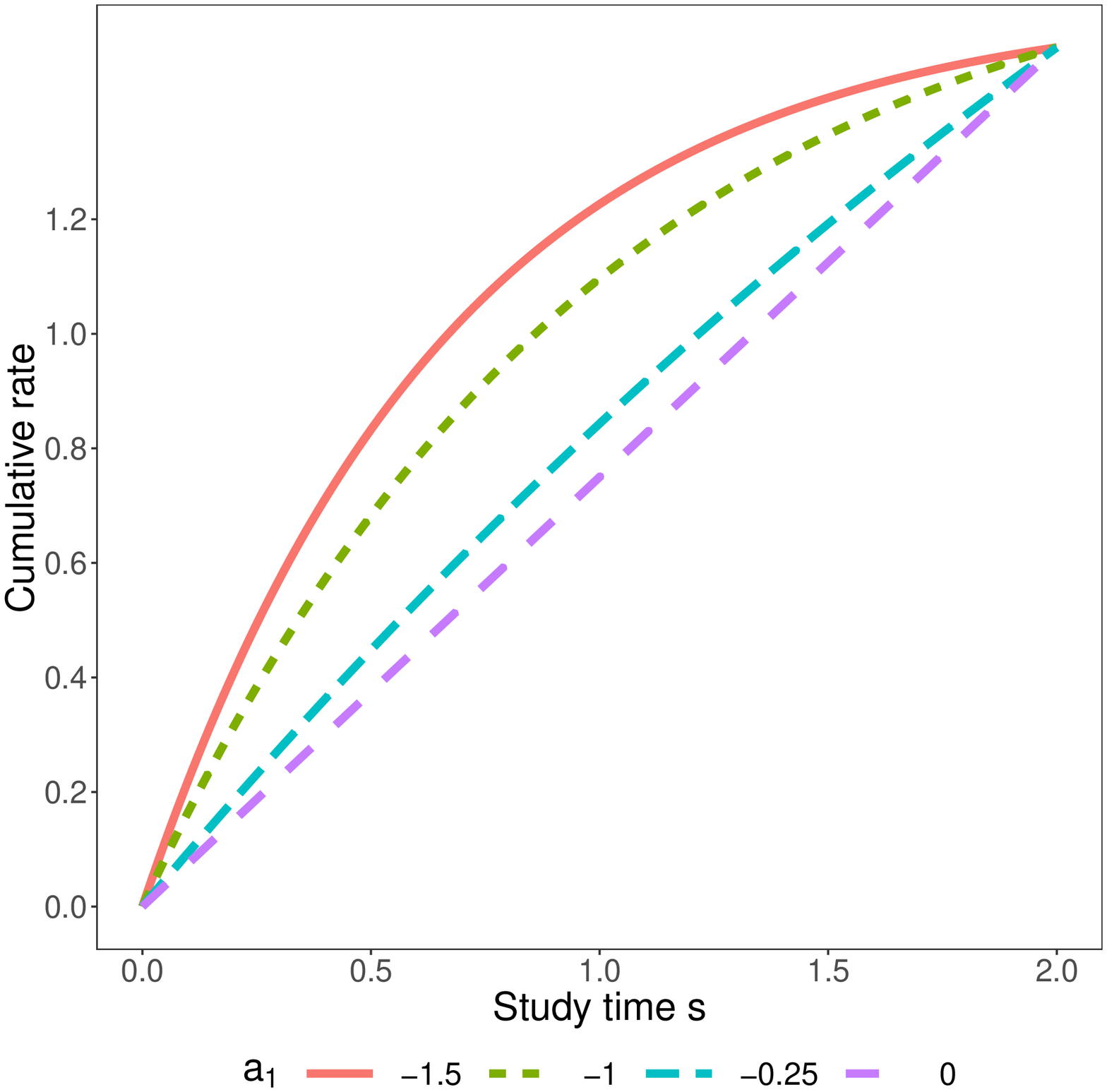}
\caption{Rate and cumulative rate for the control group for different time trends. }
\label{fig:simu_rates}
\end{figure}

The parameters for the simulation study are motivated by the clinical trial in pediatric multiple sclerosis published by  \cite{chitnis2018trial}, which was discussed in Section \ref{sec:example}.
In detail, we focus on a clinical trial with a planned maximum individual follow-up time of two years and a recruitment period of two years. 
The individual follow-up times cannot be extended beyond the initially planned maximum individual follow-up of two years.
This results in a trial duration of four years for a fixed sample design. 
The target power is $P=0.8$ for a one-sided significance level of $\alpha=0.025$.
Motivated by the observed cumulative rate presented in Section \ref{sec:example}, the cumulative rate in the control group after two years is chosen to be $\Lambda_{C}(2)=1.5$, that is an annualized control rate of 0.75.
The time trend parameter is chosen to be $\alpha_{1} = -0.25, -1, -1.5$. 
A trend of $\alpha_{1}=-1.5$ is included as an extreme case as the rate at two years is less than 0.05. 
The parameter $\alpha_{0}$ is chosen such that the cumulative rate is $1.5$ after two years.
The rate and cumulative rate functions are illustrated in Figure \ref{fig:simu_rates}.
Under the null hypothesis, the rates are equal in both groups, that is $\exp(\beta_{H_0})=1$ and under the alternative hypothesis, we assume rate ratios of $\exp(\beta_{H_1})=0.5, 0.7$. 
The shape parameter is $\phi=1.25$.
For Setting I, which describes the scenario in which the planning assumptions are correct, the sample size is chosen to be the fixed design sample size $n_{fix}$ required for a power of $P$. 
For Setting II, the sample size is $1.5n_{fix}$ to describe an overpowered clinical trial.
The fixed design sample size is $n_{fix}=148$ for $\exp(\beta_{H_1})=0.5$, and $n_{fix}=510$ for $\exp(\beta_{H_1})=0.7$. 
The trend parameter $\alpha_{1}$ does not affect the sample size in the fixed design as the fixed design is planned with an identical follow-up time for each subject.
\begin{table}\small
\caption{{\label{table:scenarios}}Scenarios considered in the simulation study motivated by the clinical trial example from Section \ref{sec:example}.}
\centering
\begin{tabular}{l*{1} {c}}
\toprule
\textbf{Parameter} &  \textbf{Value}  \\ \midrule
One-sided significance level $\alpha$ & 0.025 \\
Target power $P$ & 0.8\\
Maximum individual follow-up  [years] & 2   \\
Recruitment period [years] & 2   \\
Study duration [years] & 4  \\ 
Start time of information monitoring & After 0.5 years  \\ 
Rate ratio $\exp(\beta_{H_0})$ under $H_0$ & 1 \\
Rate ratio $\exp(\beta_{H_1})$ under $H_1$ & 0.5, 0.7 \\
Cumulative rate in control group after two years & $\Lambda_{C}(2) = 1.5$ \\
Time trend $\alpha_{1}$ & -0.25, -1, -1.5 \\
Shape parameter $\phi$ & 1.25\\
Sample size allocation $(w_{T}, w_{C})$ & $(0.5, 0.5)$   \\ \midrule
\textbf{Setting I}\\
Planned total sample size $n_{Plan}$ & $n_{fix}$ \\
\textbf{Setting II}\\
Planned total sample size $n_{Plan}$ & $1.5n_{fix}$ \\
\bottomrule
\end{tabular}
\end{table}
In addition to the blinded continuous information monitoring procedure proposed in Section \ref{sec:monitoring}, we include the blinded continuous information monitoring procedure of \cite{friede2018blinded} in our simulation study to assess its robustness concerning time trends in the rate and to compare its performance to the procedure proposed in Section \ref{sec:monitoring}. 
The monitoring procedure by \cite{friede2018blinded} will be summarized in Section \ref{sec:BCMconst}.
\subsection{Monitoring procedure by Friede et al.}
\label{sec:BCMconst}
\cite{friede2018blinded} proposed an information monitoring procedure for the negative binomial model with a constant rate $\mu_{i}$, $i=T,C$, and dispersion parameter $\varphi$, where the hypotheses of interest are also defined through the rate ratio, that is 
\begin{align*}
\tilde{H}_{0}:\frac{\mu_T}{\mu_C} \geq 1 \qquad \text{vs.} \qquad  \tilde{H}_{1}:\frac{\mu_T}{\mu_C} < 1.
\end{align*}
Here, we utilize a maximum likelihood test based on the differences of log-rates $\log(\mu_{T})-\log(\mu_{C})$ to test the null hypothesis $\tilde{H}_{0}$.
Then, the information at calendar time $t$ is given by 
\begin{align*}
\mathcal{J}_{t} = \frac{1}{\frac{1}{I_{Tt}}+\frac{1}{I_{Ct}}}, \qquad I_{it}=\sum_{j=1}^{n_i}\frac{S_{ij}^{(t)}\mu_{i}}{1+\varphi S_{ij}^{(t)}\mu_{i}}.
\end{align*}
Analogously to the information monitoring procedure proposed in Section \ref{sec:monitoring}, to monitor the information $\mathcal{J}_{t}$ while maintaining the blinding, the parameters $\mu_T$, $\mu_C$, and $\varphi$ are estimated blinded using either the lumping approach or the mixture approach. For details on the blinded parameter estimation, we refer to Section 4.1 in \cite{friede2018blinded}.
Denote the blinded estimators for the rate parameters $\mu_{i}$, $i=T,C$, and the dispersion parameter $\varphi$ by $\hat{\mu}_{(b)T}$, $\hat{\mu}_{(b)C}$, and $\hat{\varphi}_{(b)}$, respectively. 
Based on the blinded parameter estimators, the information  $\mathcal{J}_{t}$ is estimated through plug-in estimation by plugging in the following estimators:
\begin{align*}
\hat{I}_{(b)it}=w_{i}\sum_{j=1}^{n}\frac{S_{(b)j}^{(t)}\hat{\mu}_{(b)i}}{1+\hat{\varphi}_{(b)} S_{(b)j}^{(t)}\hat{\mu}_{(b)i}}, \quad i=T,C.
\end{align*}
We refer to this procedure by \textit{BCM-Const--Mix} when the blinded parameter estimation utilizes the mixture approach and  \textit{BCM-Const--Lump} when the blinded parameter estimation utilizes the lumping approach.
These procedures test the null hypothesis $\tilde{H}_{0}$ using the standard negative binomial model, that is the model without time trend. 
This is in contrast to the BCM-Trend procedures which perform the analysis based on \eqref{eq:waldstat} under the assumption of a time trend.
\subsection{Operating characteristics for Setting I}
\label{sec:simulations_set1}
In the following, we present the results of the simulation study for Setting I, that is the setting in which the planned sample size is equal to the sample size required in the fixed design to achieve the target power $P$.
The simulation results presented in this section are based on $50\,000$ Monte Carlo replications. \par
\begin{table}\small
\caption{{\label{table:setI_siglevel}}Simulated type I error rate for parameters from Table  \ref{table:scenarios} for the statistical test in the fixed design and the monitoring procedures.  Here, $\exp(\beta_{H_1})$ denotes the effect used for information monitoring. 
The Monte Carlo error is $0.0007$ for a simulated type I error rate of $0.025$.}
\centering
\begin{tabular}{{l}{l}{r}{r}{r}{r}{r}}
  \toprule
$\exp(\beta_{H_1})$ & $\alpha_1$ & Fixed & Const--Lump& Const--Mix & Trend--Lump & Trend--Mix\\ 
  \midrule
0.5 & -0.25 & 0.0272 & 0.0276 & 0.0269 & 0.0273 & 0.0267 \\ 
    & -1    & 0.0286 & 0.0285 & 0.0292 & 0.0277 & 0.0283 \\ 
    & -1.5  & 0.0278 & 0.0287 & 0.0286 & 0.0279 & 0.0277 \\ 
0.7 & -0.25 & 0.0250 & 0.0260 & 0.0254 & 0.0259 & 0.0252 \\ 
    & -1    & 0.0259 & 0.0267 & 0.0274 & 0.0264 & 0.0263 \\ 
    & -1.5  & 0.0256 & 0.0260 & 0.0260 & 0.0255 & 0.0253 \\ 
   \bottomrule
\end{tabular}
\end{table}
Table \ref{table:setI_siglevel} lists the simulated type I error rate for parameters from Table \ref{table:scenarios}.
The simulated type I error rate for the monitoring procedures generally deviate less than two times the Monte Carlo error from the simulated type I error rate of the fixed sample design.
For the scenarios with the larger effect $\exp(\beta_{H_1})=0.5$, which corresponds to the scenarios with the smaller sample size, the type I error rate is inflated for the fixed sample design and so are the type I error rates of the monitoring procedures. 
This type I error rate inflation is due to the finite sample properties of the Wald test. 
In comparison, for the scenarios with a larger sample size, that is for the effect size $\exp(\beta_{H_1})=0.7$, the type I error rate is closer to the nominal level for all procedures. 
The time trend has no noticeable effect on the type I error rates.
The simulated type I error rates of the BCM-Const procedures are larger than the simulated type I error rates of the BCM-Trend procedures.
However, the difference is not of practical relevance. 
Whether the difference between the simulated type I error rates of the BCM-Const and the BCM-Trend procedures is due to the different stopping times, see Table S1 in the supplementary material, or due to BCM-Const not explicitly accounting for the time trend is not evident from this simulation study.
Table S2 in the supplementary material shows that the continuous information monitoring does not introduce any noticeable bias in the effect estimates at the end of the trial: among all scenarios and methods, the maximum simulated bias for estimators of $\exp(\beta)=1$ is smaller than or equal to $0.0015$.\par
\begin{table}\small
\caption{\label{table:power_n}Simulated power for parameters from Table \ref{table:scenarios} for the statistical test in the fixed design and the designs with blinded continuous information monitoring procedure. The Monte Carlo error is $0.0018$ for a simulated power of $0.8$.}
\centering
\begin{tabular}{llrrrrr}
  \hline
$\exp(\beta_{H_1})$ & $\alpha_1$ & Fixed & Const--Lump & Const--Mix & Trend--Lump & Trend--Mix\\ 
  \hline
0.5 & -0.25 & 0.8095 & 0.7936 & 0.7784 & 0.7923 & 0.7764 \\ 
    & -1    & 0.8134 & 0.7996 & 0.7867 & 0.7958 & 0.7809 \\ 
    & -1.5  & 0.8086 & 0.7970 & 0.7850 & 0.7913 & 0.7749 \\ 
0.7 & -0.25 & 0.8026 & 0.7916 & 0.7873 & 0.7913 & 0.7867 \\ 
    & -1    & 0.8040 & 0.7948 & 0.7908 & 0.7910 & 0.7862 \\ 
    & -1.5  & 0.8056 & 0.7989 & 0.7952 & 0.7932 & 0.7885 \\ 
   \hline
\end{tabular}
\end{table}
Table \ref{table:power_n} shows that the blinded information monitoring procedures miss the target power by up to two percentage points. 
The monitoring procedures based on the lumping approach perform better than the procedures based on the mixture approach. 
This is due to the overestimation of the dispersion parameter in the lumping approaches which results in a later stopping times, see Table S4 in the supplementary material. 
Comparing the procedures BCM-Const and BCM-Trend for a given method of blinded parameter estimation (lumping or mixture approach), the time trends do not affect the power noticeably with differences in power of less than $0.01$.
The differences in power are associated with differences in the mean stopping time: a smaller power corresponds to an earlier mean stopping time. 
Monitoring procedures with a similar power have similar mean stopping times. 
The average stopping time of procedures based on the lumping approach is around 3.5 years which can be up to four months later compared to procedures based in the mixture approach. \par
In conclusion, when the planned sample size is identical to the sample size required in the fixed design to achieve the target power, the monitoring procedures based on the lumping approach have a power within one percentage point of the target power and can result in stopping the trial on average around six months earlier.
The monitoring procedures do not result in a reduction of the sample size.  
For designs with a continuous monitoring, the mean stopping time under the null hypothesis $H_0$ is smaller than the mean stopping time under the alternative hypothesis $H_1$. This is due to the larger overall event rate under the null hypothesis $H_0$. 
\subsection{Operating characteristics for Setting II}
\label{sec:simulations_set2}
In the following, we study the operating characteristics for the setting in which the planned sample size is $50\%$ larger that what would be required in the fixed sample design to achieve the target power. 
As before, the results are based on $50\,000$ Monte Carlo replications.
Table \ref{table:setII_siglevel} lists the simulated type I error rates.\par 
\begin{table}
\caption{\label{table:setII_siglevel}Simulated type I error rate for parameters from Table \ref{table:scenarios} for the statistical test in the fixed design and the monitoring procedures. Here, $\exp(\beta_{H_1})$ denotes the effect used for information monitoring. The Monte Carlo error is $0.0007$ for a simulated type I error rate of $0.025$.}
\centering
\begin{tabular}{llrrrrr}
\hline
$\exp(\beta_{H_1})$ & $\alpha_1$ & Fixed & Const--Lump & Const--Mix & Trend--Lump & Trend--Mix\\ 
\hline
0.5 & -0.25 & 0.0281 & 0.0281 & 0.0277 & 0.0276 & 0.0276 \\ 
    & -1    & 0.0275 & 0.0289 & 0.0287 & 0.0278 & 0.0282 \\ 
    & -1.5  & 0.0252 & 0.0279 & 0.0287 & 0.0272 & 0.0272 \\ 
0.7 & -0.25 & 0.0252 & 0.0256 & 0.0255 & 0.0250 & 0.0251 \\ 
    & -1    & 0.0248 & 0.0263 & 0.0264 & 0.0249 & 0.0256 \\ 
    & -1.5  & 0.0253 & 0.0276 & 0.0275 & 0.0260 & 0.0263 \\ 
\hline
\end{tabular}
\end{table}
For the scenarios with $\exp(\beta_{H_1})=0.5$, the type I error rates in fixed sample designs and in designs with information monitoring are on average $2.7\%$ and $2.8\%$, respectively. 
The small type I error rate inflation in these scenarios can be explained by the small sample size. 
In particular, the final sample size in the designs with information monitoring is on average smaller than in the fixed sample design by up to $30\%$, depending on the effect size, monitoring procedure, and time trend. 
For scenarios with a larger effect size, $\exp(\beta_{H_1})=0.7$, the statistical test in the fixed sample design controls the type I error rate. 
The simulated type I error rates for the BCM-Trend procedures are within a range of two times the Monte Carlo error of the simulated type I error rates in the fixed sample design.
For BCM-Const, this only holds for the time trend $\alpha_{1}=-0.25$. 
The simulated type I error rate for BCM-Const increases in the time trend $\alpha_{1}$ resulting in a noticeable type I error rate inflation. 
The magnitude of the inflation is small with about $0.2$ percentage points. \par
Table \ref{table:setII_power} shows the simulated power. 
\begin{table}
\caption{\label{table:setII_power}Simulated power for parameters from Table \ref{table:scenarios} for the statistical test in the fixed design and the monitoring procedures. The Monte Carlo error is $0.0018$ for a simulated power of $0.8$.}
\centering
\begin{tabular}{llrrrrr}
  \hline
$\exp(\beta_{H_1})$ & $\alpha_1$ & Fixed & Const--Lump & Const--Mix & Trend--Lump & Trend--Mix\\ 
  \hline
0.5 & -0.25 & 0.9333 & 0.8259 & 0.8010 & 0.8222 & 0.7979 \\ 
    & -1 & 0.9361 & 0.8389 & 0.8150 & 0.8260 & 0.8006 \\ 
    & -1.5 & 0.9348 & 0.8420 & 0.8159 & 0.8244 & 0.7964 \\ 
0.7 & -0.25 & 0.9300 & 0.8120 & 0.8048 & 0.8082 & 0.8003 \\ 
    & -1 & 0.9295 & 0.8243 & 0.8169 & 0.8097 & 0.8007 \\ 
    & -1.5 & 0.9307 & 0.8306 & 0.8228 & 0.8080 & 0.8000 \\ 
   \hline
\end{tabular}
\end{table}
Table \ref{table:setII_power} shows that the fixed design is overpowered as anticipated when increasing the required sample size by $50\%$. 
The information monitoring procedures counteract this. 
However, the BCM-Const monitoring procedures do not fully mitigate the overpowering and yield a power that is up to four percentage points larger than the target power. 
Moreover, the overpowering increases as the time trend increases for the BCM-Const procedures.
Applying the mixing approach for the blinded parameter estimation in the information monitoring results in a power closer to the target compared to applying the lumping approach.
The monitoring procedures BCM-Trend, which explicitly account for a time trend, perform better for the scenarios presented in Table \ref{table:setII_power}.
In particular, BCM-Trend--Mix achieves the target power for the considered effect sizes and time trends.
BCM-Trend--Lump is overpowered for the larger effect size.
The mean stopping time is about two years or less, see Table S10 in the supplementary material. 
Thus, the monitoring procedures result in shorter clinical trials and also reduce the sample size compared to the fixed sample design.\\
Summarizing, information monitoring can prevent overpowering of clinical trials with recurrent events and time-depending rates. 
The information monitoring procedure BCM-Const proposed by \cite{friede2018blinded} can result in overpowered and longer running clinical trial when the event rates are time-dependent. 
For the considered scenarios, the monitoring procedures mitigate the overpowering due to a too large sample size by shortening the trial duration and as such the subjects' follow-up times. 
\section{Motivating example revisited}
\label{sec:example2}
In this section, we revisit the clinical trial in pediatric multiple sclerosis, introduced in Section \ref{sec:example}. 
We illustrate the four methods for blinded continuous information monitoring discussed in this manuscript using data from the motivating example \citep*{chitnis2018trial}.
Thereto, we estimate for a calendar time during the course of the trial the information based on the data which was available at said calendar time. 
For this blinded information estimation, we assume a treatment effect of $\exp(\beta_{H_1})=0.5$, which corresponds to the planning alternative of the  actual trial, and a treatment effect of $\exp(\beta_{H_1})=0.2$, which corresponds to the treatment effect eventually observed \citep*{chitnis2018trial}.
Table \ref{table:target_info} lists the information level $\mathcal{I}_{fix}$ required for the considered effects $\exp(\beta_{H_1})$ when a power of $P=80\%$ or $P=90\%$ is targeted.\par
\begin{figure}[ht]
\centering
\includegraphics[width=0.95\textwidth]{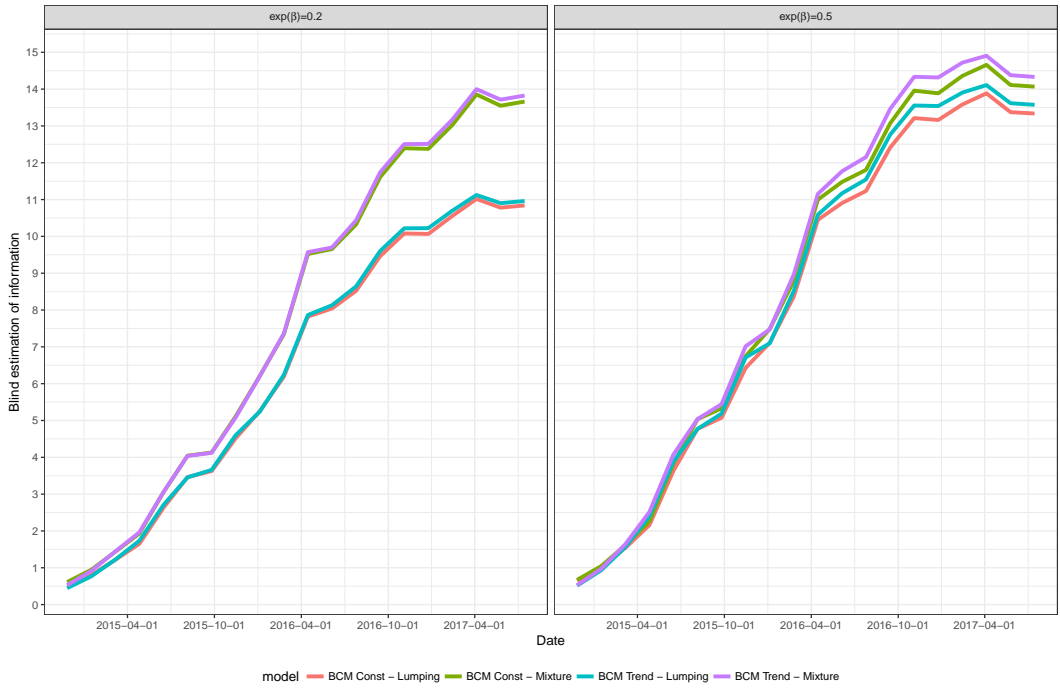}
\caption{Blinded information estimates versus the calendar time for data from a clinical trial in pediatric multiple sclerosis under assumed treatment effects of $\exp(\beta_{H_1})=0.2$ and $\exp(\beta_{H_1})=0.5$. }
\label{fig:info_time}
\end{figure}

\begin{table}
\caption{\label{table:target_info}Target information $\mathcal{I}_{fix}$ for a one-sided significance level $\alpha=2.5\%$.}
\centering
\begin{tabular}{l*{1} {c}{c}}
\toprule
\textbf{Treatment effect} $\exp(\beta_{H_1})$ & \textbf{Target power} $P$ & \textbf{Target information} $\mathcal{I}_{fix}$  \\ \midrule
0.2 & $80\%$ & 3.03\\
    & $90\%$ & 4.06\\
0.5 & $80\%$ & 16.34\\
    & $90\%$ & 21.87\\
\bottomrule
\end{tabular}
\end{table}
Figure \ref{fig:info_time} plots the blinded information estimated for four monitoring procedures versus the calendar time for data from the clinical trial in pediatric multiple sclerosis published by \cite{chitnis2018trial}.
Figure \ref{fig:info_time} shows that for an assumed effect of $\exp(\beta_{H_1})=0.2$ the information estimates from the monitoring procedures based on the lumping approach differ substantially from the information estimates of procedures using the mixture approach, in particular for dates close the end of the trial. 
Moreover, whether a monitoring procedure accounts for the time trend explicitly has no practical relevance; in other words, the procedures BCM-Trend and BCM-Const are almost identical for the lumping and the mixture approach, respectively.
Since the target information $\mathcal{I}_{fix}$ for an effect of $\exp(\beta_{H_1})=0.2$ is between 3 and 4, depending on the target power, all four monitoring procedures would have stopped within a couple of months of each other almost two years before the actual end of the trial.
For an effect of $\exp(\beta_{H_1})=0.5$, the monitoring procedures using a mixture approach result in larger information estimates than the procedures based on the lumping approach, analogous to an assumed effect of $\exp(\beta_{H_1})=0.2$. 
The difference between the monitoring procedure with the largest information estimate, BCM-Trend--Mixture, and the procedure with the smaller information estimate, BCM-Const--Lumping, is at most one. 
For the majority of the trial, this difference in the estimates has no relevant impact on the potential stopping time as the information increases by one every one to two months. 
However, towards the end of the trial, that is after October 2016 in the considered trial, the information curves become flat such that a difference of one can have a substantial effect of half a year or more. 
None of the considered methods in this section achieve the information $I_{fix}=16.34$ required for target power of $P=80\%$. 
It is important to emphasize that during the conduct of the clinical trial in pediatric multiple sclerosis, the information $I_{fix}=16.34$ was achieved. 
For the information monitoring during the clinical trial covariates were included. 
Since including covariates reduces the variability, the information is larger when covariates are included in the blinded parameter estimation.
Moreover, the planning assumption $\exp(\beta_{H_1})=0.5$ turned out to be conservative and the information required to achieve a target power of $P=80\%$ under the observed effect was reached more than 1.5 years before the end of the trial. 
\section{Discussion}
\label{sec:discussion}
Although the blinded monitoring procedure BCM-Const assuming constant rates previously proposed by \cite{friede2018blinded} turned out to be robust to some degree to time dependencies of the event rates, the procedures accounting for time trends were found to have more favorable operating characteristics. 
The proposed monitoring procedures generally did not inflate the type I error rate beyond levels observed in the fixed sample designs for the asymptotic Wald test in any practically relevant way and did not bias the treatment effect estimates in the final analysis. 
Both properties are important from a regulatory point of view. 
If the planning assumptions are correct (Setting 1), application of the information monitoring can still lead to some considerable time savings. 
When the planning assumptions were too conservative (Setting 2), the savings in terms of time and sample size are of course more pronounced. The differences between the lumping and the mixture approaches are generally small with a tendency for the lumping approach to stop trials later resulting in higher power. \par
Here we used the blinded continuous information monitoring to stop a trial early, if the information level specified in the planning had be reached. This could be due to higher event rates or less pronounced overdispersion than assumed, a scenario encountered in the trial in pediatric MS by  \cite{chitnis2018trial}. If the blinded assessment of the nuisance parameters reveals that the planning assumptions were too optimistic, then the initially planned sample size or maximum follow-up time might be increased of course. We did not explore this here as this was not relevant to our motivating example, but the combination of blinded continuous monitoring with sample size  or group-sequential design is of practical interest and will be explored by our group in the future. \par
The analyses of randomized controlled trials are often adjusted for stratification factors of the randomization or important prognostic variables. Although the procedures presented here could in principle be expanded to regression models adjusted for covariates, we did not investigate this any further. However, sample size re-estimation for covariate adjusted analyses with overdispersed count data has recently been considered by \cite{zapf2018blinded}.
As the motivating example shows, the inclusion of covariates in the analysis can lead to earlier completion of a trial.
Moreover, the procedures presented here also apply to other time trends than the log-linear trend introduced in \eqref{eq:model_trend}. For instance, one could apply period functions modeling seasonal trends which are common in asthma and chronic obstructive pulmonary disease (COPD). 
\subsection*{Acknowledgments}
  The authors wish to thank Nikolaos Sfikas for helpful comments.  

\subsection*{Author contributions}
  TM, SS, and TF conceived the concept of this study. 
  TM conducted all numerical evaluations for the examples, and drafted the manuscript.
  TM ans SS conducted the numerical evaluations for the simulations. 
  TM, SS, NB, HS, and TF critically reviewed and made substantial contributions to the manuscript. 
  All authors commented  on and approved the final manuscript.

\subsection*{Conflict of interest}
  TM ans HS are employed by Novartis Pharma AG, and own stocks thereof. TF provided consultancies to Novartis Pharma AG regarding sample size re-estimation strategies for the pediatric MS study that served as an example in this paper. 

\bibliographystyle{rss}
\bibliography{bibfile}

\begin{thebibliography}{21}
\expandafter\ifx\csname natexlab\endcsname\relax\def\natexlab#1{#1}\fi
\expandafter\ifx\csname url\endcsname\relax
  \def\url#1{\texttt{#1}}\fi
\expandafter\ifx\csname urlprefix\endcsname\relax\def\urlprefix{URL: }\fi

\bibitem[{Asendorf et~al.(2017)Asendorf, Henderson, Schmidli and
  Friede}]{asendorf2017modelling}
Asendorf, T., Henderson, R., Schmidli, H. and Friede, T. (2017) Modelling and
  sample size reestimation for longitudinal count data with incomplete follow
  up.
\newblock \textit{Statistical Methods in Medical Research}, \textbf{0}, 1--17.
\newblock \urlprefix\url{https://doi.org/10.1177/0962280217715664}.

\bibitem[{Asendorf et~al.(2018)Asendorf, Henderson, Schmidli and
  Friede}]{asendorf2018sample}
--- (2018) Sample size re-estimation for clinical trials with longitudinal
  negative binomial counts including time trends.
\newblock \textit{Statistics in Medicine}.

\bibitem[{Chitnis et~al.(2018)Chitnis, Arnold, Banwell
  et~al.}]{chitnis2018trial}
Chitnis, T., Arnold, D.~L., Banwell, B. et~al. (2018) Trial of fingolimod
  versus interferon beta-1a in pediatric multiple sclerosis.
\newblock \textit{New England Journal of Medicine}, \textbf{379}, 1017--1027.

\bibitem[{Cook(2013)}]{cook2013authors}
Cook, R.~J. (2013) Authors' redress on 'robustness of methods for blinded
  sample size re-estimation with overdispersed count data'.
\newblock \textit{Statistics in Medicine}, \textbf{32}, 3955--3957.

\bibitem[{Cook et~al.(2009)Cook, Bergeron, Boher and Liu}]{cook2009two}
Cook, R.~J., Bergeron, P.-J., Boher, J.-M. and Liu, Y. (2009) Two-stage design
  of clinical trials involving recurrent events.
\newblock \textit{Statistics in Medicine}, \textbf{28}, 2617--2638.

\bibitem[{{European Medicines Agency (EMA)}(2007)}]{EMAadaptive}
{European Medicines Agency (EMA)} (2007) Reflection paper on methodological
  issues in confirmatory clinical trials planned with an adaptive design.
\newblock
  https://www.ema.europa.eu/en/methodological-issues-confirmatory-clinical-trials-planned-adaptive-design.
\newblock 2018-12-02.

\bibitem[{{Food and Drug Administration (FDA)}(2018)}]{FDAadaptive}
{Food and Drug Administration (FDA)} (2018) Adaptive designs for clinical
  trials of drugs and biologics - guidance for industry.
\newblock
  \urlprefix\url{https://www.fda.gov/downloads/Drugs/GuidanceComplianceRegulatoryInformation/Guidances/UCM201790.pdf}.
\newblock 2018-12-02.

\bibitem[{Friede et~al.(2018)Friede, H{\"a}ring and
  Schmidli}]{friede2018blinded}
Friede, T., H{\"a}ring, D.~A. and Schmidli, H. (2018) Blinded continuous
  monitoring in clinical trials with recurrent event endpoints.
\newblock \textit{Pharmaceutical Statistics}.

\bibitem[{Friede and Kieser(2002)}]{friede2002inappropriateness}
Friede, T. and Kieser, M. (2002) On the inappropriateness of an em algorithm
  based procedure for blinded sample size re-estimation.
\newblock \textit{Statistics in Medicine}, \textbf{21}, 165--176.

\bibitem[{Friede and Miller(2012)}]{friede2012blinded}
Friede, T. and Miller, F. (2012) Blinded continuous monitoring of nuisance
  parameters in clinical trials.
\newblock \textit{Journal of the Royal Statistical Society: Series C (Applied
  Statistics)}, \textbf{61}, 601--618.

\bibitem[{Friede and Schmidli(2010{\natexlab{a}})}]{friede2010blinded1}
Friede, T. and Schmidli, H. (2010{\natexlab{a}}) Blinded sample size
  reestimation with count data: methods and applications in multiple sclerosis.
\newblock \textit{Statistics in Medicine}, \textbf{29}, 1145--1156.

\bibitem[{Friede and Schmidli(2010{\natexlab{b}})}]{friede2010blinded}
--- (2010{\natexlab{b}}) Blinded sample size reestimation with negative
  binomial counts in superiority and non-inferiority trials.
\newblock \textit{Methods of Information in Medicine}, \textbf{49}, 618--624.

\bibitem[{Jennison and Turnbull(2000)}]{jennison1999group}
Jennison, C. and Turnbull, B.~W. (2000) \textit{Group sequential methods with
  applications to clinical trials}.
\newblock Boca Raton, FL: Chapman-Hall/CRC.

\bibitem[{Lawless(1987{\natexlab{a}})}]{lawless1987negative}
Lawless, J. (1987{\natexlab{a}}) Negative binomial and mixed poisson
  regression.
\newblock \textit{Canadian Journal of Statistics}, \textbf{15}, 209--225.

\bibitem[{Lawless(1987{\natexlab{b}})}]{lawless1987regression}
--- (1987{\natexlab{b}}) Regression methods for poisson process data.
\newblock \textit{Journal of the American Statistical Association},
  \textbf{82}, 808--815.

\bibitem[{Nicholas et~al.(2012)Nicholas, Straube, Schmidli, Pfeiffer and
  Friede}]{nicholas2012time}
Nicholas, R., Straube, S., Schmidli, H., Pfeiffer, S. and Friede, T. (2012)
  Time-patterns of annualized relapse rates in randomized placebo-controlled
  clinical trials in relapsing multiple sclerosis: A systematic review and
  meta-analysis.
\newblock \textit{Multiple Sclerosis Journal}, \textbf{18}, 1290--1296.

\bibitem[{Schneider et~al.(2013{\natexlab{a}})Schneider, Schmidli and
  Friede}]{schneider2013blinded}
Schneider, S., Schmidli, H. and Friede, T. (2013{\natexlab{a}}) Blinded sample
  size re-estimation for recurrent event data with time trends.
\newblock \textit{Statistics in Medicine}, \textbf{32}, 5448--5457.

\bibitem[{Schneider et~al.(2013{\natexlab{b}})Schneider, Schmidli and
  Friede}]{schneider2013robustness}
--- (2013{\natexlab{b}}) Robustness of methods for blinded sample size
  re-estimation with overdispersed count data.
\newblock \textit{Statistics in Medicine}, \textbf{32}, 3623--3635.

\bibitem[{Tsiatis(2006)}]{tsiatis2006information}
Tsiatis, A.~A. (2006) Information-based monitoring of clinical trials.
\newblock \textit{Statistics in Medicine}, \textbf{25}, 3236--3244.

\bibitem[{Waksman(2007)}]{waksman2007assessment}
Waksman, J.~A. (2007) Assessment of the gould-shih procedure for sample size
  re-estimation.
\newblock \textit{Pharmaceutical Statistics}, \textbf{6}, 53--65.

\bibitem[{Zapf et~al.(2019)Zapf, Asendorf, Anten and {et al}}]{zapf2018blinded}
Zapf, A., Asendorf, T., Anten, C. and {et al} (2019) Blinded sample size
  reestimation for negative binomial regression with baseline adjustment.
\newblock \textit{In preparation}, \textbf{0}, 0.

\end{thebibliography}

\appendix
\section{Maximum likelihood theory}
\label{appendix:ml_theory}
For the sake of readability, we omit the index $t$.
Therefore, we denote the exposure time of subject $j$ receiving in group $i$ at calendar time $t$ by $S_{ij}$ instead of $S_{ij}^{(t)}$, and we denote the number of event up to time $S_{ij}$ by $N_{ij}$ instead of $N_{ijt}$.
\begin{align*}
\frac{\partial \log \mathcal{L}}{\partial \beta} &= 
\sum\limits_{j=1}^{n_{T}} N_{Tj} - \sum\limits_{j=1}^{n_{T}} \frac{\Lambda_T(S_{Tj})(1 + \phi N_{Tj})}{1 +  \phi\Lambda_T(S_{Tj})} \\
\frac{\partial \log \mathcal{L}}{\partial \alpha_{0}} &=  \sum_{i=T,C} \sum_{j=1}^{n_{i}} \sum_{k=1}^{N_{ij}} 1 -  \sum_{i=T,C} \sum_{j=1}^{n_{i}} \frac{\Lambda_i(S_{ij})(1 + \phi N_{ij})}{1 + \phi\Lambda_i(S_{ij})} \\
\frac{\partial \log \mathcal{L}}{\partial \alpha_{1}} &=  \sum_{i=T,C} \sum_{j=1}^{n_{i}} \sum_{k=1}^{N_{ij}} s_{ijk} -   \sum_{i=T,C} \sum_{j=1}^{n_{i}} \Lambda_i(S_{ij}) \left(
\frac{1 + \phi N_{ij}}{\alpha_1 (1 + \phi\Lambda_i(S_{ij}))} - 
\frac{(1 + \phi N_{ij})\exp(\alpha_1 S_{ij})S_{ij}}{\alpha_1(1 + \phi\Lambda_i(S_{ij})) (\exp(\alpha_1 S_{ij})-1)}\right) \\
\frac{\partial \log \mathcal{L}}{\partial \phi}  & = \sum_{i=T,C} \sum_{j=1}^{n_i}
\frac{N_{ij}}{\phi} 
- \frac{\Psi\left(N_{ij}+\phi^{-1}\right)}{\phi^{2}}
+ \frac{\Psi\left(\phi^{-1}\right)}{\phi^{2}}
- \frac{\Lambda_{i}(S_{ij})(N_{ij}+\phi^{-1})}{1+\phi \Lambda_{i}(S_{ij})}
+ \frac{\log\left(1+\phi \Lambda_{i}(S_{ij})\right)}{\phi^{2}}
\end{align*}
Here, $\Psi(x)=\Gamma^{\prime}(x)/\Gamma(x)$ is the digamma function and $\Psi^{\prime}(\cdot)$ is the first derivative of the digamma function. 
Next, we calculate the second partial derivatives. 
We start with the second partial derivatives with respect to $\phi$.
\begin{align*} 
\frac{\partial^2 \log \mathcal{L}}{\partial \beta \partial  \phi} 
&= - \sum\limits_{j=1}^{n_{T}} \frac{\Lambda_T(S_{Tj})(N_{Tj} - \Lambda_1(S_{Tj}))}{(1 + \phi\Lambda_T(S_{Tj}))^2} \\
\frac{\partial^2 \log \mathcal{L}}{\partial \alpha_{0} \partial  \phi}  
&= - \sum_{i=T,C} \sum_{j=1}^{n_{i}} \frac{\Lambda_i(S_{ij})(N_{ij} - \Lambda_i(S_{ij}) )}{(1 + \phi\Lambda_i(S_{ij}))^2}\\
\frac{\partial^2 \log \mathcal{L}}{\partial \alpha_{1} \partial  \phi} 
&= -  \sum_{i=T,C} \sum_{j=1}^{n_{i}} \Lambda_i(S_{ij}) \left(
\frac{N_{ij} - \Lambda_i(S_{ij})}{\alpha_1 (1 + \phi\Lambda_i(S_{ij}))^2} - 
\frac{(N_{ij} - \Lambda_i(S_{ij}))\exp(\alpha_1 S_{ij})S_{ij}}{\alpha_1(1 + \phi\Lambda_i(S_{ij}))^2(\exp(\alpha_1 S_{ij})-1)}\right)
\end{align*} 
With $\mathbb{E}[N_{ij}] = \Lambda_i(S_{ij})$, it follows that all second partial derivatives with respect to $\phi$ are zero. 
We continue with the remaining second partial derivatives. 
We define $H_{ij} = \exp(\alpha_0)\exp(x_i\beta_1)\exp(\alpha_1 S_{ij})S_{ij}$. 
\begin{align*}
E\left[-\frac{\partial^2 \log \mathcal{L}}{\partial \beta^{2}}\right]  
&= \sum_{j=1}^{n_{T}}\frac{\Lambda_{T}(S_{Tj})}{(1 + \phi\Lambda_{T}(S_{Tj}))}\\ 
E\left[-\frac{\partial^2 \log \mathcal{L}}{\partial \alpha_{0}^2}\right]  
&= \sum_{i=T,C} \sum_{j=1}^{n_{i}}\frac{\Lambda_{i}(S_{ij})}{(1 + \phi\Lambda_{i}(S_{ij}))} \\
E\left[-\frac{\partial^2 \log \mathcal{L}}{\partial \beta \partial \alpha_{0}}\right] 
&= \sum_{j=1}^{n_{T}}\frac{\Lambda_{T}(S_{Tj})}{(1 + \phi\Lambda_{T}(S_{Tj}))}\\
E\left[-\frac{\partial^2 \log \mathcal{L}}{\partial \alpha_{0} \partial \alpha_{1} }\right]
&= \sum_{i=T,C}\sum\limits_{j=1}^{n_i}\frac{H_{ij} - \Lambda_i(S_{ij})}{\alpha_1(1 + \phi\Lambda_i(S_{ij}))} \\
E\left[- \frac{\partial^2 \log \mathcal{L}}{\partial \beta \partial \alpha_1}  \right] &= \sum\limits_{j=1}^{n_T}\frac{H_{Tj} - \Lambda_T(S_{Tj})}{\alpha_1(1 + \phi\Lambda_T(S_{Tj}))} \\
E\left[- \frac{\partial^2 \log \mathcal{L}}{\partial \alpha_1^2} \right] &=  \sum_{i=T,C} \sum\limits_{j=1}^{n_i} \alpha^{-1}_{1}H_{ij}S_{ij} - \frac{\phi\left(H_{ij}^2 - (\Lambda_i(S_{ij}))^2\right) + 2\left(H_{ij} - \Lambda_i(S_{ij})\right)}{\alpha_1^2(1 + \phi\Lambda_i(S_{ij}))} \\\\            
\end{align*}
It follows that the Fisher information matrix $I$ for the parameter vector $(\alpha_{0}, \alpha_{1}, \beta, \phi)$ is given by 
\begin{align*}
\mathbf{I} = \begin{pmatrix}
  \mathbf{I}_{3x3} & 0\\ 
  0 &a\\
\end{pmatrix}
\end{align*}
with 
\begin{align*}
 I_{3x3} = 
 \mathbb{E}\left[-
\begin{pmatrix}
\frac{\partial^2 \log \mathcal{L}}{\partial \alpha_{0}^{2}} & \frac{\partial^2 \log \mathcal{L}}{\partial \alpha_{1} \partial \alpha_{0}}  & \frac{\partial^2 \log \mathcal{L}}{\partial \beta \partial \alpha_{0}} \\
\frac{\partial^2 \log \mathcal{L}}{\partial \alpha_{1} \partial \alpha_{0}} & \frac{\partial^2 \log \mathcal{L}}{\partial \alpha_{1}^{2}}  & \frac{\partial^2 \log \mathcal{L}}{\partial \beta \partial \alpha_{1}} \\
\frac{\partial^2 \log \mathcal{L}}{\partial \beta \partial \alpha_{0}} & \frac{\partial^2 \log \mathcal{L}}{\partial \beta \partial \alpha_{1}}  & \frac{\partial^2 \log \mathcal{L}}{\partial \beta^{2} } 
\end{pmatrix}\right]
\end{align*}
and 
\begin{align*}
a=  E\left[-\frac{\partial^2 \log\mathcal{L}}{\partial\phi^2}\right].
\end{align*}
To calculate the asymptotic variance of the maximum likelihood estimator for the parameter vector $(\alpha_{0}, \alpha_{1}, \beta, \phi)$, the inverse Fisher information matrix has to be calculated. 
It is given by
\begin{align*}
\mathbf{I}^{-1} = 
\begin{pmatrix}
  \mathbf{I}_{3x3}^{-1} & 0\\ 
  0 &\frac{1}{a}\\
\end{pmatrix}.
\end{align*}
The inverse $\mathbf{I}_{3x3}^{-1} $ can be calculated using the general formula for the inverse of a $3x3$-matrix.
Lastly, we calculate the second derivative with respect to $\phi$:
\begin{align*}
   \frac{\partial^{2} \log \mathcal{L}}{\partial \phi^{2}}   =  \sum_{i=T,C} \sum_{j=1}^{n_i} &
- \frac{N_{ij}}{\phi^{2}} 
+ 2\frac{\Psi\left(N_{ij}+\phi^{-1}\right)}{\phi^{3}}
+ \frac{\Psi^{\prime}\left(N_{ij}+\phi^{-1}\right)}{\phi^{4}}
- 2 \frac{\Psi\left(\phi^{-1}\right)}{\phi^{3}}
-  \frac{\Psi^{\prime}\left(\phi^{-1}\right)}{\phi^{4}}\\
&+  \frac{\Lambda_{i}(S_{ij}) \phi^{-2} (1+\phi \Lambda_{i}(S_{ij})) + \Lambda_{i}(S_{ij})^2(N_{ij}+1/\phi) }{\left(1+\phi \Lambda_{i}(S_{ij})\right)^{2}} \\
& +  \frac{\Lambda_{i}(S_{ij}) \phi - 2\log(1+\phi \Lambda_{i}(S_{ij}))(1+\phi \Lambda_{i}(S_{ij}))}{\phi^{3}(1+\phi \Lambda_{i}(S_{ij}))}.
\end{align*}
The expected value of the second derivative of the log-likelihood with respect to $\phi$ has no closed form expression.

\end{document}